\documentclass[twocolumn,showpacs,amsmath,prl,aps,amssymb,superscriptaddress,nofootinbib]{revtex4-1} 
\usepackage[T1]{fontenc}
\usepackage[latin9]{inputenc}
\usepackage{times}
\usepackage{color} 
\usepackage{xspace}
\usepackage{amssymb,amsmath}
\usepackage{amsbsy}
\usepackage[pdftex]{graphicx}
\usepackage{bm}
\usepackage{float}
\usepackage{adjustbox}
\usepackage{epstopdf}

\usepackage[unicode,breaklinks]{hyperref}
\hypersetup{
    unicode=true,
    plainpages=false, 
    colorlinks=true,
    linkcolor=blue,
    citecolor=blue,
    filecolor=black,
    urlcolor=blue
}
\usepackage{url}
\usepackage{verbatim}

\begin{document}

\title{Coarsening dynamics of an isotropic ferromagnetic superfluid}
\author{Lewis A. Williamson}  
\affiliation{Dodd-Walls Centre for Photonic and Quantum Technologies, Department of Physics, University of Otago, Dunedin 9016, New Zealand}
\author{P.~B.~Blakie}  
\affiliation{Dodd-Walls Centre for Photonic and Quantum Technologies, Department of Physics, University of Otago, Dunedin 9016, New Zealand}
\affiliation{Swinburne University of Technology, Sarawak Campus, School of Engineering, Computing and Science, Jalan Simpang Tiga, 93350 Kuching, Sarawak, Malaysia}

\begin{abstract}
In zero magnetic field the ground state manifold of a ferromagnetic spin-1 condensate is SO(3) and exhibits $\mathbb{Z}_2$ vortices as topological defects. We investigate the phase ordering dynamics of this system after being quenched into this ferromagnetic phase from a zero temperature unmagnetized phase.  Following the quench, we observe the ordering of both magnetic and gauge domains. We find that these domains grow diffusively, i.e.~with domain size $L(t)\sim t^{1/2}$, and exhibit dynamic scale invariance. The coarsening dynamics progresses as $\mathbb{Z}_2$ vortices annihilate, however we find that at finite energy a number of these vortices persist in small clumps without influencing magnetic or gauge order. We consider the influence of a small non-zero magnetic field, which reduces the ground state symmetry, and show that this sets a critical length scale such that when the domains reach this size the system dynamically transitions in order parameter and scaling behaviour from an isotropic  to  an anisotropic ferromagnetic superfluid.
\end{abstract}

\maketitle

When a many-body system is quenched across a critical point into a symmetry breaking phase, the formation of order in the new phase may undergo universal dynamics~\cite{Bray1994}. During or soon after the quench, causally disconnected spatial domains of the order parameter develop, with each domain making an independent choice of the symmetry breaking phase. The choice of phases define an equilibrium state manifold. The domains then grow and compete for the global equilibrium state. Often the system exhibits dynamic scale invariance when the size of the domains $L(t)$ is larger than microscopic length scales. The domains then grow as $L(t)\sim t^{1/z}$ where $z$ is the dynamic critical exponent that defines the dynamic universality class of the system~\cite{Bray1994}. The dynamic universality class is determined by general properties of the system, such as symmetries and conservation laws, and the nature of the equilibrium state manifold. The nature of the equilibrium state manifold also determines what topological defects are supported, and the annihilation of such defects play an integral role in the coarsening dynamics~\cite{Bray1994}.

Much of the previous work on coarsening dynamics has focussed on phenomenological dissipative models in classical systems. Bose-Einstein condensates provide an isolated quantum system with a tractable microscopic description to explore coarsening dynamics. Furthermore, condensates have gauge symmetries that when combined with spin symmetries offer a rich array of ground state manifolds and topological defects to explore. Symmetry breaking has been observed following quantum phase transitions in ferromagnetic~\cite{Sadler2006a,Leslie2009a,Guzman2011a} and antiferromagnetic~\cite{Bookjans2011b,kang2017} spin-1 condensates, as well as in immiscible binary condensates~\cite{De2014a}. Simulations of phase ordering in the easy-axis and easy-plane phases in ferromagnetic spin-1 condensates~\cite{Kudo2013a,williamson2016a,williamson2016b} and in an immiscible binary condensate~\cite{hofmann2014} reveal phase ordering consistent with classical dynamic universality classes~\cite{Hohenberg1977}. Coarsening dynamics in systems with $\text{U}(1)$ gauge invariance has also been explored~\cite{Damle1996a,karl2017,bourges2017,kulczykowski2017}, however this manifold is shared by a classical $\text{XY}$ spin system. Exploring coarsening dynamics in quantum systems with more complex manifolds arising from gauge invariance has so far received little attention.

\begin{figure}
\centering
\includegraphics[trim=15cm 18cm 11cm 37cm, clip=true,width=0.45\textwidth]{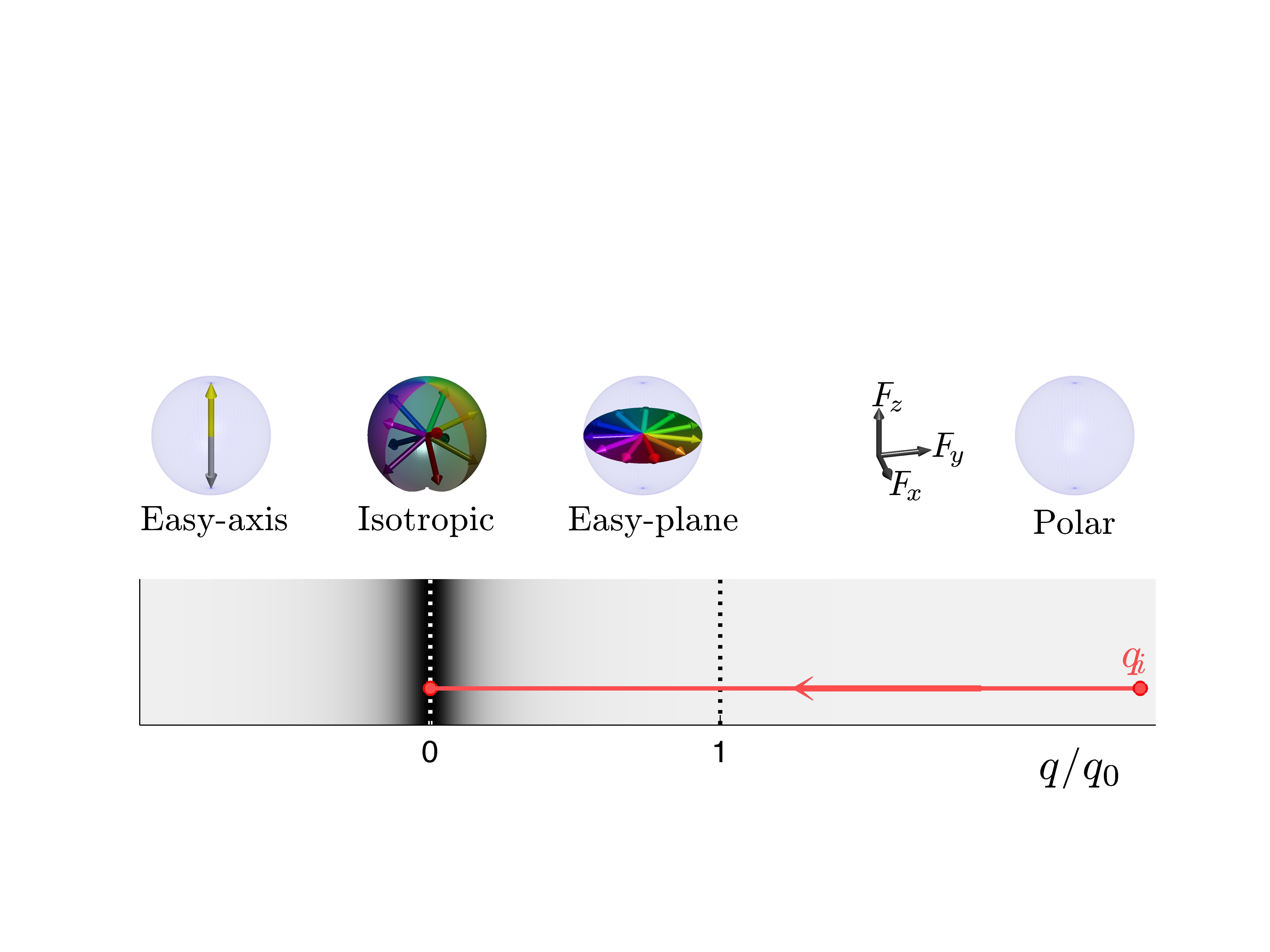}
\caption{\label{phaseDiag}Ground state phase diagram of a ferromagnetic spin-1 condensate as a function of the quadratic Zeeman energy $q$. The net $F_z$ magnetization is assumed to be 0. For large $q$, the condensate is unmagnetized (polar phase). For $q<q_0\equiv 2|g_s|n_0$, the system magnetizes. The direction of magnetization is: along $z$ for $q<0$, termed easy-axis; isotropic for $q=0$; and in the $xy$-plane for $0<q<q_0$, termed easy-plane. The choice of spin directions in each phase is shown by arrows in the respective spin spheres. The red arrow indicates an instantaneous quench from deep in the polar phase to $q=0$.}
\end{figure}

The ground state of an isolated ferromagnetic condensate exhibits an $\text{SO}(3)$ symmetry, arising from the full spin and gauge invariance of the system. Symmetry breaking in this phase has been observed in the spin-1 case following a temperature quench~\cite{Guzman2011a}. A quantum quench from the unmagnetised polar phase to the $\text{SO}(3)$ phase can also be induced by quenching the quadratic Zeeman field, see Fig.~\ref{phaseDiag}. Coarsening dynamics in an $\text{SO}(3)$ system has recently been explored in an antiferromagnetic lattice system, but such systems exhibit geometrical frustration that suppresses the coarsening~\cite{phuc2017}. Apart from this, coarsening dynamics in an $\text{SO}(3)$ system has to our knowledge not been explored. An $\text{SO}(3)$ system supports $\mathbb{Z}_2$ vortices that we can expect to be present during the coarsening dynamics. Theoretical studies have shown that $\mathbb{Z}_2$ vortices can be stablised in rotating condensates~\cite{isoshima2002,mizushima2002,lovegrove2012}, however much less work has explored their role in dynamical processes. In addition, questions regarding a topological phase transition between bound and unbound $\mathbb{Z}_2$ vortices have drawn interest in frustrated antiferromagnetic lattices~\cite{kawamura1984,kawamura2010}, but so far results are inconclusive.

In this letter we explore coarsening dynamics in the $\text{SO}(3)$ phase of a ferromagnetic spin-1 condensate following a quench from the polar phase to the isotropic phase. The quench is implemented by a sudden change in $q$ from a large positive energy to $q=0$, see Fig.~\ref{phaseDiag}. We find that coarsening of both spin and gauge domains occur, and that both ordering processes exhibit dynamic scale invariance with a critical exponent of $z=2$. We find that the quench generates many $\mathbb{Z}_2$ vortices that initially decay in a way consistent with the formation of order, but for longer times decay much slower with a persistence of closely ``bound'' vortices. We also explore how the coarsening dynamics behaves for small but non-zero $|q|$ and identify a transition between two different dynamic universality classes that occurs when the domains grow larger than a critical size.  Our results indicate that it should be possible to explore the dynamic transition to anisotropic scaling in current experiments.

\paragraph{System.} We consider a quasi-two-dimensional spin-1 condensate described by the energy functional \cite{Ho1998a,Ohmi1998a}
\begin{align}\label{spinH}
H\!=\!\int\!d^2\bm{x}\left[\bm{\psi}^\dagger\!\left(\!-\!\frac{\hbar^2\nabla^2}{2M}\!+\!qf_z^2\right)\!\bm{\psi}+\frac{g_n}{2}n^2\!+\!\frac{g_s}{2}\left|\bm{F}\right|^2\!\right]\!,
\end{align}
where $\bm{\psi}=\left(\psi_1,\psi_0,\psi_{-1}\right)^T$ is the spin-1 field, $n=\bm{\psi}^\dagger\bm{\psi}$ is the total areal density, and $\bm{F}$ is the spin density with components $F_\mu=\bm{\psi}^\dagger f_\mu\bm{\psi}$, where $f_\mu\in\{f_x,f_y,f_z\}$ are the spin-1 matrices. The interactions are parameterized by the density dependent ($g_n$) and spin-dependent ($g_s$) coupling constants. For stability we must have $g_n>0$. Ferromagnetic interactions occur when $g_s<0$, i.e.~the system maximizes $|\mathbf{F}|$.  These conditions are realized for $^{87}$Rb atoms in the ground state $F=1$ hyperfine manifold   \cite{Chang2004a}. A magnetic field along $z$  shifts the energies of the spin states. We neglect the linear Zeeman shift in (\ref{spinH}) since it can be removed by transforming $\bm{\psi}$ into a frame rotating at the Larmor frequency. The quadratic Zeeman shift $q$  plays a crucial role in determining the ground state, and can be tuned independently of the magnetic field using external microwave fields (e.g.~see \cite{Gerbier2006a}). 

The ground state magnetic phases for varying $q$ are shown in Fig.~\ref{phaseDiag}. Here we are interested in the phase ordering near $q=0$. The ground state spinor at $q=0$ can be written as~\cite{Ho1998a}
\begin{align}\label{gstate}
\bm{\psi}=\sqrt{n_0}e^{i\theta}\left(\begin{array}{c}e^{-i\phi}\cos^2\frac{\beta}{2}\\\frac{1}{\sqrt{2}}\sin\beta\\e^{i\phi}\sin^2\frac{\beta}{2}\end{array}\right),
\end{align}
where $\theta$ is the phase of the spinor component $\psi_0$ and $\{\phi,\beta\}$ determine the direction of spin density, $\bm{F}=n_0(\sin\beta\cos\phi,\sin\beta\sin\phi,\cos\beta)$. Spatial variation of $\theta$ gives rise to $\psi_0$ gauge domains, while spatial variation of $\phi$ and $\beta$ gives rise to magnetic domains.

The full symmetry of the ground state~\eqref{gstate} is $\text{SO}(3)$~\cite{Ho1998a}. The first homotopy group of $\text{SO}(3)$ is $\mathbb{Z}_2$, so that the state~\eqref{gstate} exhibits two distinct topologies, one of which is the defect free state~\cite{Ho1998a}. Therefore only singly charged defects exist, called $\mathbb{Z}_2$ vortices. The manifold $\text{SO}(3)$ is diffeomorphic to the real projective space $\text{RP}^3$, which is the 3-sphere $S^3$ with antipodal points identified. A $\mathbb{Z}_2$ vortex occurs when the order parameter around a loop in the physical system maps onto a path joining two antipodal points in $\text{S}^3$, which forms a loop in $\text{RP}^3$ that cannot be continuously undone. The annihilation of topological defects is intimately linked with phase ordering~\cite{Bray1994}. We expect $\mathbb{Z}_2$ vortices to be produced by the quench and decay during the coarsening dynamics. The $\mathbb{Z}_2$ vortices couple the gauge and spin ordering, as a circulation in the gauge angle $\theta$ can be continuously transformed into a circulation in the spin angles $\phi$ so that gauge defects and spin defects are not independent~\cite{Ho1998a,Kawaguchi2012R}. This is in contrast to the easy-axis phase, where the gauge and magnetic degrees of freedom support distinct defects and the ordering of each is different~\cite{bourges2017}.  

\paragraph{Coarsening dynamics.}
To simulate the quench dynamics we numerically evolve the spin-1 Gross-Pitaevskii equations (GPEs) \cite{Kawaguchi2012R} on a $1024\times1024$ grid with initial condition of a  polar condensate $\bm{\psi}=\sqrt{n_0}(0,1,0)^T$ that has vacuum noise added to Bogoliubov modes with $q_i=\infty$ ($q_i$ is the initial quadratic Zeeman energy) according to the truncated Wigner prescription \cite{williamson2016b}. The noise is necessary to seed the formation of symmetry breaking domains. We study the growth and coarsening of the order parameters $\varphi=\left\{\bm{F}/n_0,\psi_0/\sqrt{n_0}\right\}$ by examining the correlation functions
\begin{align}\label{GF}
G_\varphi(r,t)=&\left\langle\varphi(\bm{r})^\dagger\varphi(\bm{0})\right\rangle_t-\left|\left\langle\varphi(\bm{0})\right\rangle_t\right|^2,
\end{align}
where the average is taken at a time $t$ after the quench. We perform the average by invoking spatial and rotational invariance of correlations, and also average over four simulation trajectories conducted with different initial noise. We use a  condensate density of  $n_0=10^4/\xi_s^2$, where $\xi_s\equiv\hbar/\sqrt{q_0 M}$ is the spin healing length. The system size is $l\times l=800\xi_s\times 800\xi_s$ and $g_n/|g_s|=10$. Microscopic details such as coupling strengths and grid resolution affect microscopic dynamics but not late time universal scaling~\cite{williamson2016b,bourges2017}.

\begin{figure}
\centering
\includegraphics[width=0.5\textwidth]{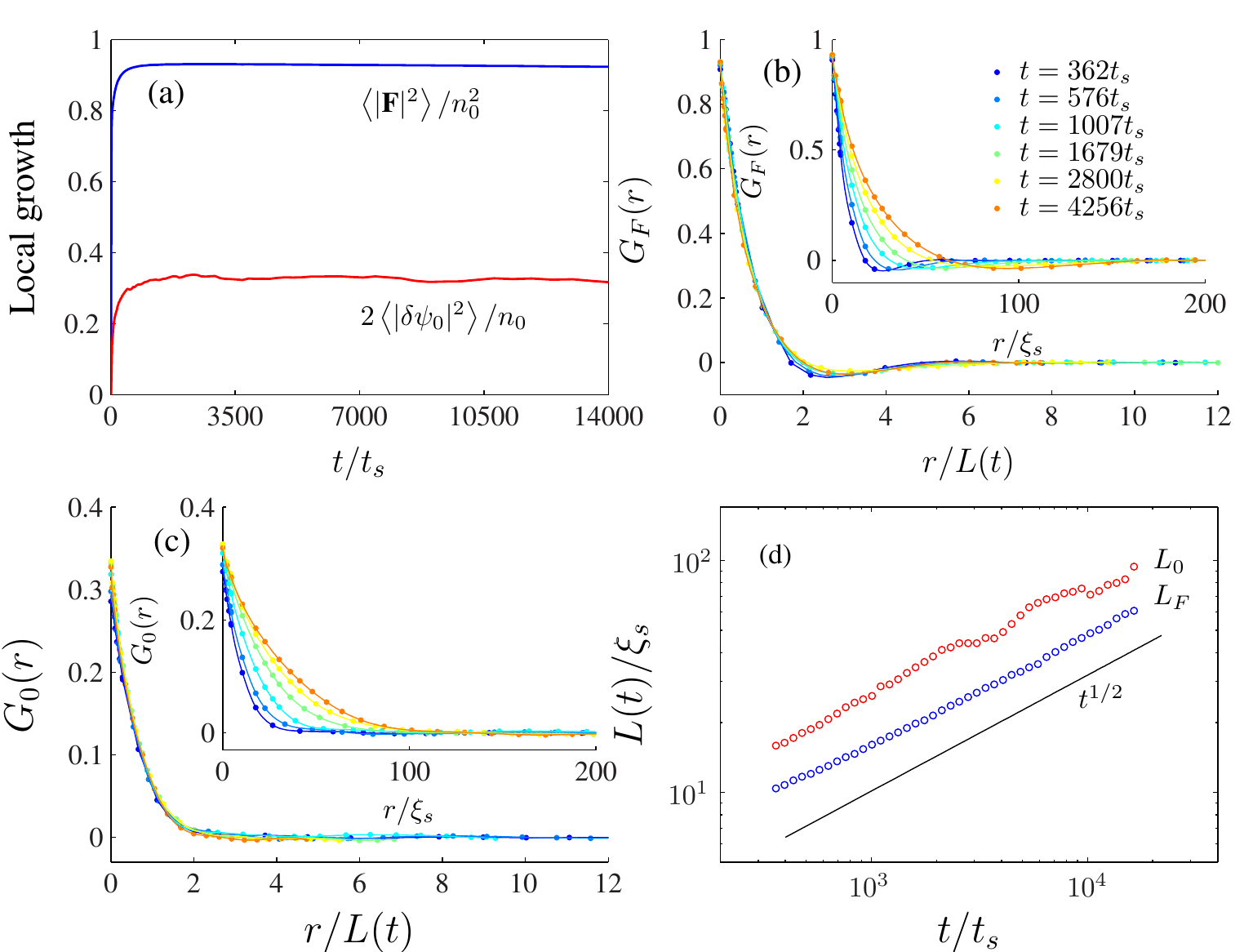}
\caption{\label{GLfig}(a) Growth of mean squared magnetization and $\psi_0$ fluctuations $\delta\psi_0\equiv\psi_0-\left\langle\psi_0\right\rangle$. Both quantities grow until $t\sim 500 t_s$, and then remain  steady.
The final value of mean squared magnetization is less than the ground state value $n_0^2$ due to heating from the quench.
The spatial correlation functions of (b) $\bm{F}/n_0$ and (c)  $\psi_0/\sqrt{n_0}$ [see Eq.~\eqref{GF}] during the coarsening dynamics. Insets:  correlation functions for various times versus the spatial coordinate $r$. Main plots: correlation functions for various times versus the rescaled coordinates $r/L_F(t)$ for (b) and $r/L_0(t)$ for (c).
Correlation function collapse onto a single curve shows the domain growth exhibits dynamic scale invariance. (d) The evolution of $L_F(t)$ and $L_0(t)$ versus time, showing that they grow as $t^{1/2}$.
}
\end{figure}

At $t=0$ the quadratic Zeeman shift is instantaneously quenched to zero. The initial exponential growth of unstable modes resulting in a growth of the spin density $\mathbf{F}$ is well understood~~\cite{Saito2007a,Leslie2009a,Barnett2011,zhang2005}. For times $t\gg t_s\equiv \hbar/q_0$ this growth stabilises, see Fig.~\ref{GLfig}(a). Across these long evolution times coherent domains of spin and $\psi_0$ fluctuations coarsen.

We examine the growth of domains by plotting the correlation functions~\eqref{GF} for various times, see Fig.~\ref{GLfig}(b),(c). The spatial extent of correlations grow with time (insets) while the shape of the correlation function is preserved. This shows that the domain growth is scale invariant, consistent with the theory of phase ordering dynamics. Rescaling the position coordinate of the correlation functions $G_\varphi$  by the respective domain size  $L_\varphi(t)$  results in the correlation functions at different times collapsing onto a single curve, where we take $L_\varphi(t)$  to be the shortest distance where $G_\varphi(r,t)=0.2G_\varphi(0,t)$. We plot the growing length scales $L_F(t)$ and $L_0(t)$ in Fig.~\ref{GLfig}(d). We see that the domains grow as a power law with time $\sim t^{1/z}$, and a fit to this growth gives $z=2$. For long wavelength (low energy) dynamics, the spin denstiy $\bm{F}$ obeys a Landau-Lifshitz equation (LLE) modified due to the advection of the order parameter by the superfluid velocity~\cite{lamacraft2008}. Imposing that the LLE equation obeys dynamic scale invariance gives a dynamic critical exponent $z=(d+2)/2$ for a $d$ dimensional system at low temperatures~\cite{Tauber2014} (see also~\cite{kudo2016}), consistent with our results.

\paragraph{$\mathbb{Z}_2$ vortices.} In addition to long wavelength excitations, the state~\eqref{gstate} supports topological defects known as $\mathbb{Z}_2$ vortices. The nontrivial topology of $\mathbb{Z}_2$ vortices comes about as follows. The superfluid velocity for the state~\eqref{gstate} is $M\bm{v}/\hbar=\nabla\theta-\nabla\phi+\left(1-\cos\beta\right)\nabla\phi$ where single-valuedness of the state imposes that $\oint_Cd\theta=2\pi n_\theta$, $\oint_Cd\phi=2\pi n_\phi$ and $\oint_Cd\beta=0$ for a closed path $C$, where $n_\theta$ and $n_\phi$ are integers, and $\left(1-\cos\beta\right)\nabla\phi$ is the Berry phase~\cite{Kawaguchi2012R}. The difference between the superfluid velocity and Berry phase is $\nabla(\theta-\phi)$, which is quantized. A state with even phase winding of $\nabla(\theta-\phi)$, i.e.\ even $n_\theta-n_\phi$, is homotopic to the vortex free state, while all odd phase windings are topologically equivalent to a single phase winding~\cite{Ho1998a,Kawaguchi2012R}. The group of homotopically distinct topologies in the $\text{SO}(3)$ system is therefore isomorphic to the group $\mathbb{Z}_2$, hence the vortex name.

\begin{figure}
\centering
\includegraphics[width=3.0in]{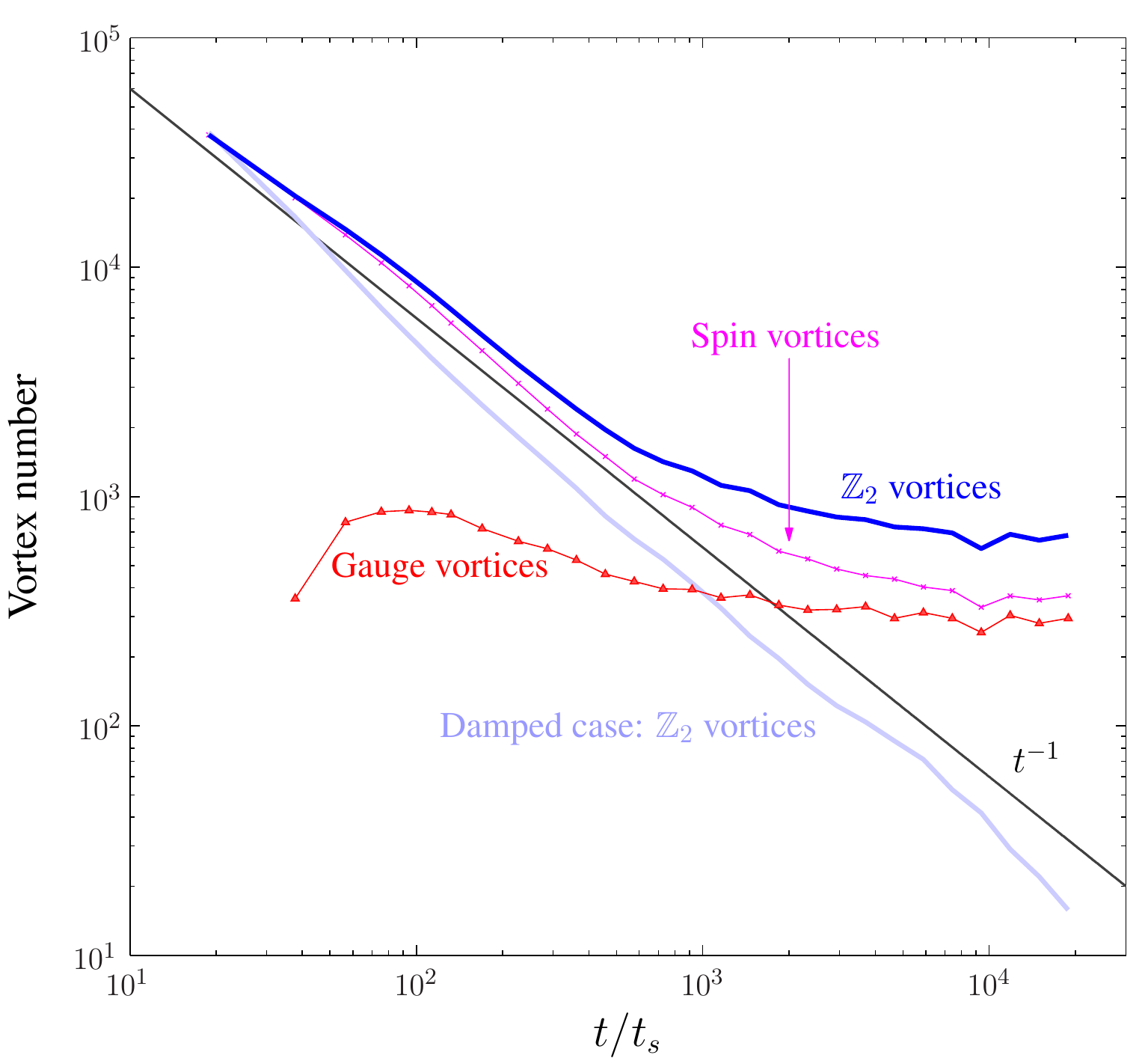}
\caption{\label{Z2num}Decay of the number of $\mathbb{Z}_2$ vortices (which can be subdivided into  ``spin'' and ``gauge'' vortices) with time. Results are the average of four simulations.  For $t\lesssim 10^3t_s$, the number of $\mathbb{Z}_2$ vortices decays as $t^{-1}$, consistent with magnetic and gauge ordering. During this time the number of spin vortices is much larger than the number of gauge vortices. For $t>10^3t_s$, the decay of $\mathbb{Z}_2$ vortices decreases substantially leading to a significant number of residual ``bound'' vortices. Results for the number of  $\mathbb{Z}_2$ vortices for a damped GPE simulation are also shown. 
 }
\end{figure}

The positions of $\mathbb{Z}_2$ vortices can be identified by points around which the phase winding of $\theta-\phi$ is odd. Figure~\ref{Z2num} shows the decay of the total number of $\mathbb{Z}_2$ vortices versus time. Vortex annihilation occurs through the collision of two oddly charged vortices. For $t\lesssim 10^3t_s$, the number of $\mathbb{Z}_2$ vortices decays as $t^{-1}$. This decay is consistent with the average distance between vortices growing as $t^{1/2}$, i.e.~the same scaling as the domain growth. Thus the annihilation of vortices appears associated with the coarsening dynamics. For $t>10^3t_s$, the decay of $\mathbb{Z}_2$ vortices decreases substantially, even though the magnetic and gauge domains continue to grow. The residual $\mathbb{Z}_2$ vortices must therefore be present in small (``bound'') groups  with a net even charge, so as not to destroy the magnetic or gauge order. We expect these bound vortices arise out of heating from the quench. We have also carried out simulations using the damped spin-1 GPEs~\cite{Su2011a,Rooney2012a,Bradley2014a}, which remove excess thermal energy and drives the system toward the vortex free ground state. In this case there is a continual $t^{-1}$ decay of $\mathbb{Z}_2$ vortices [see Fig.~\ref{Z2num}], consistent with magnetic and gauge ordering (note that we obtain a $t^{1/2}$ growth of magnetic and gauge domains in the damped regime). The bound $\mathbb{Z}_2$ vortices therefore annihilate when damping is included, as expected.

The phase winding of a $\mathbb{Z}_2$ vortex is made up of $\theta$ and $\phi$ phase windings. We detect  $\mathbb{Z}_2$ vortices by looking for such phase winding around plaquettes of size twice our simulation grid spacing. We find these occur predominantly as  ``spin vortices'' with $n_\phi=\pm 1$, $n_\theta=0$ and ``gauge vortices'' with $n_\theta=\pm 1$, $n_\phi=0$. It is not true, however, that spin vortices affect only spin order and gauge vortices affect only gauge order. Figure~\ref{Z2num} shows that throughout the simulation the number of spin vortices (at the length scale $\approx \!\! 1.6\xi_s$ of our vortex detection) is always much larger than the number of gauge vortices, even though the spin and gauge coarsening dynamics occur in sync (see Fig.~\ref{GLfig}). A vortex may consist of spin circulation at one length scale (say close to the core) but then be continuously transformed to a gauge vortex at larger length scales, and vice versa, so that spin and gauge vortices detected at one length scale can each affect both spin and gauge order at another length scale.

\paragraph{Coarsening for non zero $|q|$.} During the coarsening for $q=0$, all of the components of $\bm{F}$ exhibit dynamic scale invariance with a length scale that grows as $t^{1/2}$. In comparison, for the easy-axis phase with $q<0$ only the $F_z$ correlations exhibit scale invariant growth but with a $t^{2/3}$ growth law, while for the easy-plane phase with $0<q<q_0$ only the $F_\perp\equiv F_x+iF_y$ correlations exhibit scale invariant growth with a  $t/\log t$ growth law~\cite{williamson2016a,williamson2016b}. This raises the interesting question of how the scaling of correlations of $F_z$ and $F_\perp$ change as $|q|\rightarrow 0$, pertinent to spinor experiments which are able to resolve $q$ to uncertainties of $\delta q\lesssim10^{-2}q_0$ \cite{Vinit2017a}.

\begin{figure}
\includegraphics[width=0.5\textwidth]{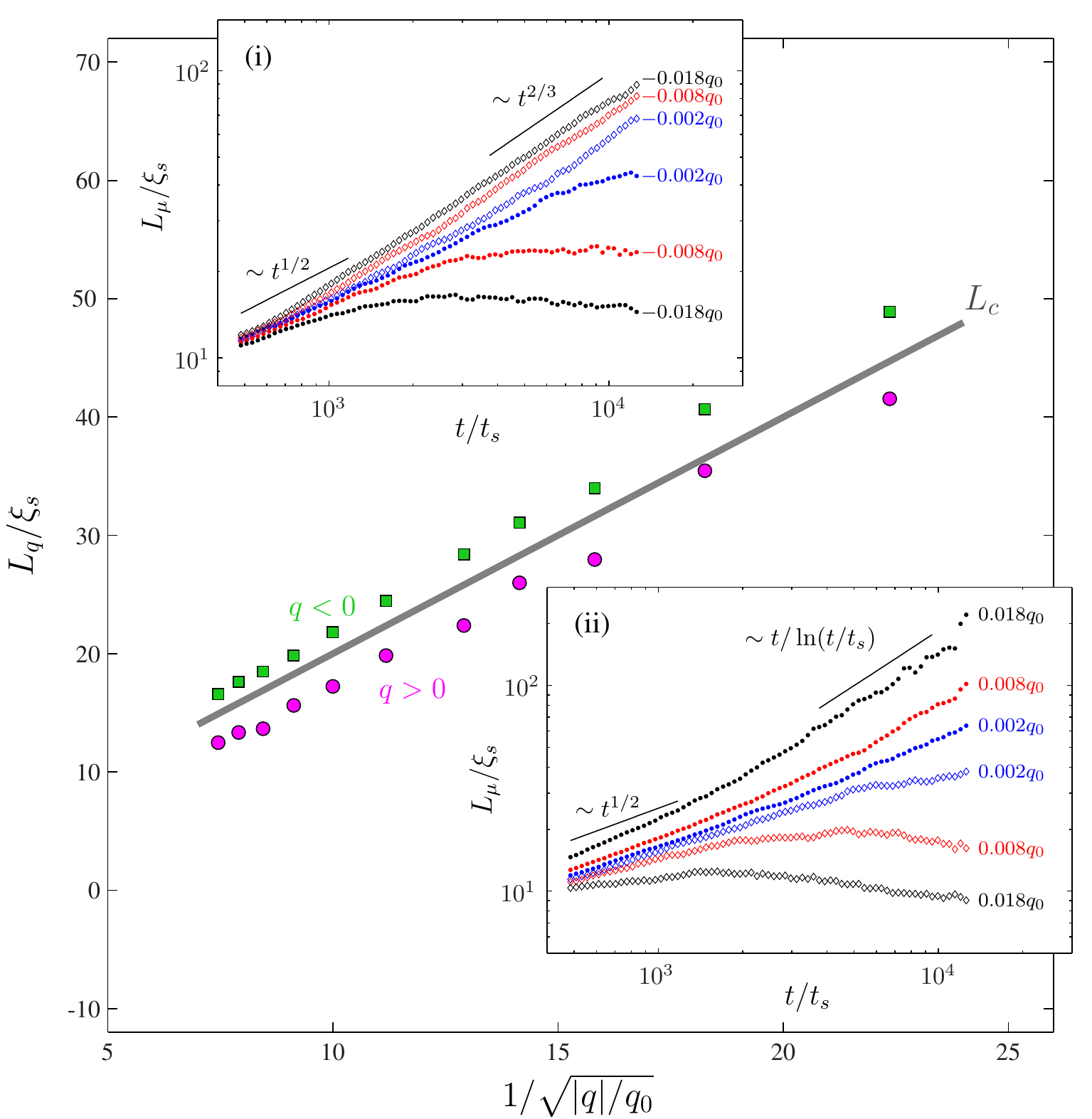}
\caption{\label{smallq} The domain size $L_q$ at which coarsening transitions from isotropic to anisotropic ordering, for $q>0$   (magenta circles)  and $q<0$  (green squares). We take $L_q=\max_t L_z$  $(L_q=\max_t  L_\perp )$ for   $q>0$  $(q<0)$. The line shows $L_c$ (see text). Insets show the growth of $F_\perp$ (filled circles) and $F_z$ (unfilled diamonds) domains versus $t$ for    (i)  $q<0$ and (ii)  $q>0$  (as labelled in insets). 
}
\end{figure}

We study finite $q$ effects using the correlation functions
\begin{align}\label{Gmu}
G_\mu(r,t)=\frac{1}{n_0^2}\left\langle F^*_\mu(\bm{r})F_\mu(\bm{0})\right\rangle_t,
\end{align}
where $\mu=z,\perp$, and extract domain sizes $L_\mu$ for different times by finding the shortest distance at which $G_\mu(r,t)=0.2 G_\mu(0,t)$. The insets to Fig.~\ref{smallq} show plots of the domain sizes $L_\mu$ versus time for quenches to a range of small $q$ values.  
 Initially the domains grow as $t^{1/2}$, like for $q=0$. As the average domain size increases
we observe that the system transitions to a new dynamic universality class: for $q<0$ [Inset (i)] $F_\perp$ correlations stop growing and then begin to shrink, while the easy-axis order ($F_z$) transitions to growing as $t^{2/3}$; for $q>0$ [Inset (ii)] $F_z$ correlations stop growing and then begin to shrink, while the easy-plane order ($F_\perp$) transitions to growing as $t/\log t$.  Phase ordering with a growth law that depends on length scale also occurs in classical binary fluids~\cite{furukawa1985,Bray1994}, however what we observe here differs in that the order parameter also changes during the transition.
 
To quantify the transition from isotropic to anisotropic coarsening we identify the transition length $L_q$ as the maximum domain size for the non-ordering spin component. That is,  $L_q$ is the maximum value of $L_z$ ($L_\perp$) for $q>0$ ($q<0$). The transition length $L_q$ is plotted in Fig.~\ref{smallq} where we observe that for positive and negative $q$ values $L_q$ is well described by a critical length  $L_c=2\hbar/\sqrt{M|q|}$, obtained by equating the domain kinetic energy to $|q|$.

\paragraph{Conclusion.} In this work we have explored the magnetic and gauge phase ordering dynamics of a ferromagnetic spin-1 condensate quenched to the isotropic phase. We have found that the growth of domains is scale invariant with a dynamic critical exponent of $z=2$. We identify $\mathbb{Z}_2$ vortices that annihilate as order grows. We also observe a dynamic phase transition from isotropic to anisotropic magnetic ordering for quenches to experimentally achievable small values of $|q|$. It would be interesting to consider this dynamic transition from the perspective of non-thermal fixed points, which have been argued to be more general than equilibrium critical points~\cite{nowak2012,langen2016,karl2017}. Another interesting future direction arising from this work is to compare the $\text{SO}(3)$ ordering observed here with ordering for an $\text{S}^3$ order parameter, which does not support topologically protected $\mathbb{Z}_2$ vortices. This ground state manifold occurs at the miscibility transition in a binary condensate~\cite{karl2013}. It would also be interesting to explore if the phase ordering observed here progresses to a state with quasi-long range order or to a state where order decays exponentially~\cite{mukerjee2006}. This could shed light on the existence of a topological phase transition between bound and unbound $\mathbb{Z}_2$ vortices. We note that while ground state properties of $\mathbb{Z}_2$ vortices have been explored~\cite{isoshima2002,mizushima2002,lovegrove2012}, we are not aware of studies of $\mathbb{Z}_2$ vortex dynamics. Such a study may be possible using variational Lagrangian methods employed for polar-core and half-quantum spin vortices~\cite{williamson2016c,kasamatsu2016}.

\begin{acknowledgements}
We acknowledge valuable discussions with Y. Kawaguchi, T. Simula, I. Spielman, and X. Yu. We gratefully acknowledge support from the Marsden Fund of the Royal Society of New Zealand.
\end{acknowledgements}


\begin{thebibliography}{44}%
\makeatletter
\providecommand \@ifxundefined [1]{%
 \@ifx{#1\undefined}
}%
\providecommand \@ifnum [1]{%
 \ifnum #1\expandafter \@firstoftwo
 \else \expandafter \@secondoftwo
 \fi
}%
\providecommand \@ifx [1]{%
 \ifx #1\expandafter \@firstoftwo
 \else \expandafter \@secondoftwo
 \fi
}%
\providecommand \natexlab [1]{#1}%
\providecommand \enquote  [1]{``#1''}%
\providecommand \bibnamefont  [1]{#1}%
\providecommand \bibfnamefont [1]{#1}%
\providecommand \citenamefont [1]{#1}%
\providecommand \href@noop [0]{\@secondoftwo}%
\providecommand \href [0]{\begingroup \@sanitize@url \@href}%
\providecommand \@href[1]{\@@startlink{#1}\@@href}%
\providecommand \@@href[1]{\endgroup#1\@@endlink}%
\providecommand \@sanitize@url [0]{\catcode `\\12\catcode `\$12\catcode
  `\&12\catcode `\#12\catcode `\^12\catcode `\_12\catcode `\%12\relax}%
\providecommand \@@startlink[1]{}%
\providecommand \@@endlink[0]{}%
\providecommand \url  [0]{\begingroup\@sanitize@url \@url }%
\providecommand \@url [1]{\endgroup\@href {#1}{\urlprefix }}%
\providecommand \urlprefix  [0]{URL }%
\providecommand \Eprint [0]{\href }%
\providecommand \doibase [0]{http://dx.doi.org/}%
\providecommand \selectlanguage [0]{\@gobble}%
\providecommand \bibinfo  [0]{\@secondoftwo}%
\providecommand \bibfield  [0]{\@secondoftwo}%
\providecommand \translation [1]{[#1]}%
\providecommand \BibitemOpen [0]{}%
\providecommand \bibitemStop [0]{}%
\providecommand \bibitemNoStop [0]{.\EOS\space}%
\providecommand \EOS [0]{\spacefactor3000\relax}%
\providecommand \BibitemShut  [1]{\csname bibitem#1\endcsname}%
\let\auto@bib@innerbib\@empty
\bibitem [{\citenamefont {Bray}(1994)}]{Bray1994}%
  \BibitemOpen
  \bibfield  {author} {\bibinfo {author} {\bibfnamefont {A.}~\bibnamefont
  {Bray}},\ }\href {\doibase 10.1080/00018739400101505} {\bibfield  {journal}
  {\bibinfo  {journal} {Advances in Physics}\ }\textbf {\bibinfo {volume}
  {43}},\ \bibinfo {pages} {357} (\bibinfo {year} {1994})}\BibitemShut
  {NoStop}%
\bibitem [{\citenamefont {Sadler}\ \emph {et~al.}(2006)\citenamefont {Sadler},
  \citenamefont {Higbie}, \citenamefont {Leslie}, \citenamefont
  {Vengalattore},\ and\ \citenamefont {Stamper-Kurn}}]{Sadler2006a}%
  \BibitemOpen
  \bibfield  {author} {\bibinfo {author} {\bibfnamefont {L.~E.}\ \bibnamefont
  {Sadler}}, \bibinfo {author} {\bibfnamefont {J.~M.}\ \bibnamefont {Higbie}},
  \bibinfo {author} {\bibfnamefont {S.~R.}\ \bibnamefont {Leslie}}, \bibinfo
  {author} {\bibfnamefont {M.}~\bibnamefont {Vengalattore}}, \ and\ \bibinfo
  {author} {\bibfnamefont {D.~M.}\ \bibnamefont {Stamper-Kurn}},\ }\href
  {http://dx.doi.org/10.1038/nature05094} {\bibfield  {journal} {\bibinfo
  {journal} {Nature}\ }\textbf {\bibinfo {volume} {443}},\ \bibinfo {pages}
  {312} (\bibinfo {year} {2006})}\BibitemShut {NoStop}%
\bibitem [{\citenamefont {Leslie}\ \emph {et~al.}(2009)\citenamefont {Leslie},
  \citenamefont {Guzman}, \citenamefont {Vengalattore}, \citenamefont {Sau},
  \citenamefont {Cohen},\ and\ \citenamefont {Stamper-Kurn}}]{Leslie2009a}%
  \BibitemOpen
  \bibfield  {author} {\bibinfo {author} {\bibfnamefont {S.~R.}\ \bibnamefont
  {Leslie}}, \bibinfo {author} {\bibfnamefont {J.}~\bibnamefont {Guzman}},
  \bibinfo {author} {\bibfnamefont {M.}~\bibnamefont {Vengalattore}}, \bibinfo
  {author} {\bibfnamefont {J.~D.}\ \bibnamefont {Sau}}, \bibinfo {author}
  {\bibfnamefont {M.~L.}\ \bibnamefont {Cohen}}, \ and\ \bibinfo {author}
  {\bibfnamefont {D.~M.}\ \bibnamefont {Stamper-Kurn}},\ }\href {\doibase
  10.1103/PhysRevA.79.043631} {\bibfield  {journal} {\bibinfo  {journal} {Phys.
  Rev. A}\ }\textbf {\bibinfo {volume} {79}},\ \bibinfo {pages} {043631}
  (\bibinfo {year} {2009})}\BibitemShut {NoStop}%
\bibitem [{\citenamefont {Guzman}\ \emph {et~al.}(2011)\citenamefont {Guzman},
  \citenamefont {Jo}, \citenamefont {Wenz}, \citenamefont {Murch},
  \citenamefont {Thomas},\ and\ \citenamefont {Stamper-Kurn}}]{Guzman2011a}%
  \BibitemOpen
  \bibfield  {author} {\bibinfo {author} {\bibfnamefont {J.}~\bibnamefont
  {Guzman}}, \bibinfo {author} {\bibfnamefont {G.-B.}\ \bibnamefont {Jo}},
  \bibinfo {author} {\bibfnamefont {A.~N.}\ \bibnamefont {Wenz}}, \bibinfo
  {author} {\bibfnamefont {K.~W.}\ \bibnamefont {Murch}}, \bibinfo {author}
  {\bibfnamefont {C.~K.}\ \bibnamefont {Thomas}}, \ and\ \bibinfo {author}
  {\bibfnamefont {D.~M.}\ \bibnamefont {Stamper-Kurn}},\ }\href {\doibase
  10.1103/PhysRevA.84.063625} {\bibfield  {journal} {\bibinfo  {journal} {Phys.
  Rev. A}\ }\textbf {\bibinfo {volume} {84}},\ \bibinfo {pages} {063625}
  (\bibinfo {year} {2011})}\BibitemShut {NoStop}%
\bibitem [{\citenamefont {Bookjans}\ \emph {et~al.}(2011)\citenamefont
  {Bookjans}, \citenamefont {Vinit},\ and\ \citenamefont
  {Raman}}]{Bookjans2011b}%
  \BibitemOpen
  \bibfield  {author} {\bibinfo {author} {\bibfnamefont {E.~M.}\ \bibnamefont
  {Bookjans}}, \bibinfo {author} {\bibfnamefont {A.}~\bibnamefont {Vinit}}, \
  and\ \bibinfo {author} {\bibfnamefont {C.}~\bibnamefont {Raman}},\ }\href
  {\doibase 10.1103/PhysRevLett.107.195306} {\bibfield  {journal} {\bibinfo
  {journal} {Phys. Rev. Lett.}\ }\textbf {\bibinfo {volume} {107}},\ \bibinfo
  {pages} {195306} (\bibinfo {year} {2011})}\BibitemShut {NoStop}%
\bibitem [{\citenamefont {Kang}\ \emph {et~al.}(2017)\citenamefont {Kang},
  \citenamefont {Seo}, \citenamefont {Kim},\ and\ \citenamefont
  {Shin}}]{kang2017}%
  \BibitemOpen
  \bibfield  {author} {\bibinfo {author} {\bibfnamefont {S.}~\bibnamefont
  {Kang}}, \bibinfo {author} {\bibfnamefont {S.~W.}\ \bibnamefont {Seo}},
  \bibinfo {author} {\bibfnamefont {J.~H.}\ \bibnamefont {Kim}}, \ and\
  \bibinfo {author} {\bibfnamefont {Y.}~\bibnamefont {Shin}},\ }\href {\doibase
  10.1103/PhysRevA.95.053638} {\bibfield  {journal} {\bibinfo  {journal} {Phys.
  Rev. A}\ }\textbf {\bibinfo {volume} {95}},\ \bibinfo {pages} {053638}
  (\bibinfo {year} {2017})}\BibitemShut {NoStop}%
\bibitem [{\citenamefont {De}\ \emph {et~al.}(2014)\citenamefont {De},
  \citenamefont {Campbell}, \citenamefont {Price}, \citenamefont {Putra},
  \citenamefont {Anderson},\ and\ \citenamefont {Spielman}}]{De2014a}%
  \BibitemOpen
  \bibfield  {author} {\bibinfo {author} {\bibfnamefont {S.}~\bibnamefont
  {De}}, \bibinfo {author} {\bibfnamefont {D.~L.}\ \bibnamefont {Campbell}},
  \bibinfo {author} {\bibfnamefont {R.~M.}\ \bibnamefont {Price}}, \bibinfo
  {author} {\bibfnamefont {A.}~\bibnamefont {Putra}}, \bibinfo {author}
  {\bibfnamefont {B.~M.}\ \bibnamefont {Anderson}}, \ and\ \bibinfo {author}
  {\bibfnamefont {I.~B.}\ \bibnamefont {Spielman}},\ }\href {\doibase
  10.1103/PhysRevA.89.033631} {\bibfield  {journal} {\bibinfo  {journal} {Phys.
  Rev. A}\ }\textbf {\bibinfo {volume} {89}},\ \bibinfo {pages} {033631}
  (\bibinfo {year} {2014})}\BibitemShut {NoStop}%
\bibitem [{\citenamefont {Kudo}\ and\ \citenamefont
  {Kawaguchi}(2013)}]{Kudo2013a}%
  \BibitemOpen
  \bibfield  {author} {\bibinfo {author} {\bibfnamefont {K.}~\bibnamefont
  {Kudo}}\ and\ \bibinfo {author} {\bibfnamefont {Y.}~\bibnamefont
  {Kawaguchi}},\ }\href {\doibase 10.1103/PhysRevA.88.013630} {\bibfield
  {journal} {\bibinfo  {journal} {Phys. Rev. A}\ }\textbf {\bibinfo {volume}
  {88}},\ \bibinfo {pages} {013630} (\bibinfo {year} {2013})}\BibitemShut
  {NoStop}%
\bibitem [{\citenamefont {Williamson}\ and\ \citenamefont
  {Blakie}(2016{\natexlab{a}})}]{williamson2016a}%
  \BibitemOpen
  \bibfield  {author} {\bibinfo {author} {\bibfnamefont {L.~A.}\ \bibnamefont
  {Williamson}}\ and\ \bibinfo {author} {\bibfnamefont {P.~B.}\ \bibnamefont
  {Blakie}},\ }\href {\doibase 10.1103/PhysRevLett.116.025301} {\bibfield
  {journal} {\bibinfo  {journal} {Phys. Rev. Lett.}\ }\textbf {\bibinfo
  {volume} {116}},\ \bibinfo {pages} {025301} (\bibinfo {year}
  {2016}{\natexlab{a}})}\BibitemShut {NoStop}%
\bibitem [{\citenamefont {Williamson}\ and\ \citenamefont
  {Blakie}(2016{\natexlab{b}})}]{williamson2016b}%
  \BibitemOpen
  \bibfield  {author} {\bibinfo {author} {\bibfnamefont {L.~A.}\ \bibnamefont
  {Williamson}}\ and\ \bibinfo {author} {\bibfnamefont {P.~B.}\ \bibnamefont
  {Blakie}},\ }\href {\doibase 10.1103/PhysRevA.94.023608} {\bibfield
  {journal} {\bibinfo  {journal} {Phys. Rev. A}\ }\textbf {\bibinfo {volume}
  {94}},\ \bibinfo {pages} {023608} (\bibinfo {year}
  {2016}{\natexlab{b}})}\BibitemShut {NoStop}%
\bibitem [{\citenamefont {Hofmann}\ \emph {et~al.}(2014)\citenamefont
  {Hofmann}, \citenamefont {Natu},\ and\ \citenamefont
  {Das~Sarma}}]{hofmann2014}%
  \BibitemOpen
  \bibfield  {author} {\bibinfo {author} {\bibfnamefont {J.}~\bibnamefont
  {Hofmann}}, \bibinfo {author} {\bibfnamefont {S.~S.}\ \bibnamefont {Natu}}, \
  and\ \bibinfo {author} {\bibfnamefont {S.}~\bibnamefont {Das~Sarma}},\ }\href
  {\doibase 10.1103/PhysRevLett.113.095702} {\bibfield  {journal} {\bibinfo
  {journal} {Phys. Rev. Lett.}\ }\textbf {\bibinfo {volume} {113}},\ \bibinfo
  {pages} {095702} (\bibinfo {year} {2014})}\BibitemShut {NoStop}%
\bibitem [{\citenamefont {Hohenberg}\ and\ \citenamefont
  {Halperin}(1977)}]{Hohenberg1977}%
  \BibitemOpen
  \bibfield  {author} {\bibinfo {author} {\bibfnamefont {P.~C.}\ \bibnamefont
  {Hohenberg}}\ and\ \bibinfo {author} {\bibfnamefont {B.~I.}\ \bibnamefont
  {Halperin}},\ }\href {\doibase 10.1103/RevModPhys.49.435} {\bibfield
  {journal} {\bibinfo  {journal} {Rev. Mod. Phys.}\ }\textbf {\bibinfo {volume}
  {49}},\ \bibinfo {pages} {435} (\bibinfo {year} {1977})}\BibitemShut
  {NoStop}%
\bibitem [{\citenamefont {Damle}\ \emph {et~al.}(1996)\citenamefont {Damle},
  \citenamefont {Majumdar},\ and\ \citenamefont {Sachdev}}]{Damle1996a}%
  \BibitemOpen
  \bibfield  {author} {\bibinfo {author} {\bibfnamefont {K.}~\bibnamefont
  {Damle}}, \bibinfo {author} {\bibfnamefont {S.~N.}\ \bibnamefont {Majumdar}},
  \ and\ \bibinfo {author} {\bibfnamefont {S.}~\bibnamefont {Sachdev}},\ }\href
  {\doibase 10.1103/PhysRevA.54.5037} {\bibfield  {journal} {\bibinfo
  {journal} {Phys. Rev. A}\ }\textbf {\bibinfo {volume} {54}},\ \bibinfo
  {pages} {5037} (\bibinfo {year} {1996})}\BibitemShut {NoStop}%
\bibitem [{\citenamefont {Karl}\ and\ \citenamefont
  {Gasenzer}(2017)}]{karl2017}%
  \BibitemOpen
  \bibfield  {author} {\bibinfo {author} {\bibfnamefont {M.}~\bibnamefont
  {Karl}}\ and\ \bibinfo {author} {\bibfnamefont {T.}~\bibnamefont
  {Gasenzer}},\ }\href {\doibase https://doi.org/10.1088/1367-2630/aa7eeb}
  {\bibfield  {journal} {\bibinfo  {journal} {New J. Phys.}\ }\textbf {\bibinfo
  {volume} {19}},\ \bibinfo {pages} {093014} (\bibinfo {year}
  {2017})}\BibitemShut {NoStop}%
\bibitem [{\citenamefont {Bourges}\ and\ \citenamefont
  {Blakie}(2017)}]{bourges2017}%
  \BibitemOpen
  \bibfield  {author} {\bibinfo {author} {\bibfnamefont {A.}~\bibnamefont
  {Bourges}}\ and\ \bibinfo {author} {\bibfnamefont {P.~B.}\ \bibnamefont
  {Blakie}},\ }\href {\doibase 10.1103/PhysRevA.95.023616} {\bibfield
  {journal} {\bibinfo  {journal} {Phys. Rev. A}\ }\textbf {\bibinfo {volume}
  {95}},\ \bibinfo {pages} {023616} (\bibinfo {year} {2017})}\BibitemShut
  {NoStop}%
\bibitem [{\citenamefont {Kulczykowski}\ and\ \citenamefont
  {Matuszewski}(2017)}]{kulczykowski2017}%
  \BibitemOpen
  \bibfield  {author} {\bibinfo {author} {\bibfnamefont {M.}~\bibnamefont
  {Kulczykowski}}\ and\ \bibinfo {author} {\bibfnamefont {M.}~\bibnamefont
  {Matuszewski}},\ }\href {\doibase 10.1103/PhysRevB.95.075306} {\bibfield
  {journal} {\bibinfo  {journal} {Phys. Rev. B}\ }\textbf {\bibinfo {volume}
  {95}},\ \bibinfo {pages} {075306} (\bibinfo {year} {2017})}\BibitemShut
  {NoStop}%
\bibitem [{\citenamefont {Phuc}\ \emph {et~al.}(2017)\citenamefont {Phuc},
  \citenamefont {Momoi}, \citenamefont {Furukawa}, \citenamefont {Kawaguchi},
  \citenamefont {Fukuhara},\ and\ \citenamefont {Ueda}}]{phuc2017}%
  \BibitemOpen
  \bibfield  {author} {\bibinfo {author} {\bibfnamefont {N.~T.}\ \bibnamefont
  {Phuc}}, \bibinfo {author} {\bibfnamefont {T.}~\bibnamefont {Momoi}},
  \bibinfo {author} {\bibfnamefont {S.}~\bibnamefont {Furukawa}}, \bibinfo
  {author} {\bibfnamefont {Y.}~\bibnamefont {Kawaguchi}}, \bibinfo {author}
  {\bibfnamefont {T.}~\bibnamefont {Fukuhara}}, \ and\ \bibinfo {author}
  {\bibfnamefont {M.}~\bibnamefont {Ueda}},\ }\href {\doibase
  10.1103/PhysRevA.95.013620} {\bibfield  {journal} {\bibinfo  {journal} {Phys.
  Rev. A}\ }\textbf {\bibinfo {volume} {95}},\ \bibinfo {pages} {013620}
  (\bibinfo {year} {2017})}\BibitemShut {NoStop}%
\bibitem [{\citenamefont {Isoshima}\ and\ \citenamefont
  {Machida}(2002)}]{isoshima2002}%
  \BibitemOpen
  \bibfield  {author} {\bibinfo {author} {\bibfnamefont {T.}~\bibnamefont
  {Isoshima}}\ and\ \bibinfo {author} {\bibfnamefont {K.}~\bibnamefont
  {Machida}},\ }\href {\doibase 10.1103/PhysRevA.66.023602} {\bibfield
  {journal} {\bibinfo  {journal} {Phys. Rev. A}\ }\textbf {\bibinfo {volume}
  {66}},\ \bibinfo {pages} {023602} (\bibinfo {year} {2002})}\BibitemShut
  {NoStop}%
\bibitem [{\citenamefont {Mizushima}\ \emph {et~al.}(2002)\citenamefont
  {Mizushima}, \citenamefont {Machida},\ and\ \citenamefont
  {Kita}}]{mizushima2002}%
  \BibitemOpen
  \bibfield  {author} {\bibinfo {author} {\bibfnamefont {T.}~\bibnamefont
  {Mizushima}}, \bibinfo {author} {\bibfnamefont {K.}~\bibnamefont {Machida}},
  \ and\ \bibinfo {author} {\bibfnamefont {T.}~\bibnamefont {Kita}},\ }\href
  {\doibase 10.1103/PhysRevA.66.053610} {\bibfield  {journal} {\bibinfo
  {journal} {Phys. Rev. A}\ }\textbf {\bibinfo {volume} {66}},\ \bibinfo
  {pages} {053610} (\bibinfo {year} {2002})}\BibitemShut {NoStop}%
\bibitem [{\citenamefont {Lovegrove}\ \emph {et~al.}(2012)\citenamefont
  {Lovegrove}, \citenamefont {Borgh},\ and\ \citenamefont
  {Ruostekoski}}]{lovegrove2012}%
  \BibitemOpen
  \bibfield  {author} {\bibinfo {author} {\bibfnamefont {J.}~\bibnamefont
  {Lovegrove}}, \bibinfo {author} {\bibfnamefont {M.~O.}\ \bibnamefont
  {Borgh}}, \ and\ \bibinfo {author} {\bibfnamefont {J.}~\bibnamefont
  {Ruostekoski}},\ }\href {\doibase 10.1103/PhysRevA.86.013613} {\bibfield
  {journal} {\bibinfo  {journal} {Phys. Rev. A}\ }\textbf {\bibinfo {volume}
  {86}},\ \bibinfo {pages} {013613} (\bibinfo {year} {2012})}\BibitemShut
  {NoStop}%
\bibitem [{\citenamefont {Kawamura}\ and\ \citenamefont
  {Miyashita}(1984)}]{kawamura1984}%
  \BibitemOpen
  \bibfield  {author} {\bibinfo {author} {\bibfnamefont {H.}~\bibnamefont
  {Kawamura}}\ and\ \bibinfo {author} {\bibfnamefont {S.}~\bibnamefont
  {Miyashita}},\ }\href {\doibase 10.1143/JPSJ.53.4138} {\bibfield  {journal}
  {\bibinfo  {journal} {Journal of the Physical Society of Japan}\ }\textbf
  {\bibinfo {volume} {53}},\ \bibinfo {pages} {4138} (\bibinfo {year}
  {1984})}\BibitemShut {NoStop}%
\bibitem [{\citenamefont {Kawamura}\ \emph {et~al.}(2010)\citenamefont
  {Kawamura}, \citenamefont {Yamamoto},\ and\ \citenamefont
  {Okubo}}]{kawamura2010}%
  \BibitemOpen
  \bibfield  {author} {\bibinfo {author} {\bibfnamefont {H.}~\bibnamefont
  {Kawamura}}, \bibinfo {author} {\bibfnamefont {A.}~\bibnamefont {Yamamoto}},
  \ and\ \bibinfo {author} {\bibfnamefont {T.}~\bibnamefont {Okubo}},\ }\href
  {\doibase 10.1143/JPSJ.79.023701} {\bibfield  {journal} {\bibinfo  {journal}
  {Journal of the Physical Society of Japan}\ }\textbf {\bibinfo {volume}
  {79}},\ \bibinfo {pages} {023701} (\bibinfo {year} {2010})}\BibitemShut
  {NoStop}%
\bibitem [{\citenamefont {Ho}(1998)}]{Ho1998a}%
  \BibitemOpen
  \bibfield  {author} {\bibinfo {author} {\bibfnamefont {T.-L.}\ \bibnamefont
  {Ho}},\ }\href {\doibase 10.1103/PhysRevLett.81.742} {\bibfield  {journal}
  {\bibinfo  {journal} {Phys. Rev. Lett.}\ }\textbf {\bibinfo {volume} {81}},\
  \bibinfo {pages} {742} (\bibinfo {year} {1998})}\BibitemShut {NoStop}%
\bibitem [{\citenamefont {Ohmi}\ and\ \citenamefont
  {Machida}(1998)}]{Ohmi1998a}%
  \BibitemOpen
  \bibfield  {author} {\bibinfo {author} {\bibfnamefont {T.}~\bibnamefont
  {Ohmi}}\ and\ \bibinfo {author} {\bibfnamefont {K.}~\bibnamefont {Machida}},\
  }\href {\doibase 10.1143/JPSJ.67.1822} {\bibfield  {journal} {\bibinfo
  {journal} {J. Phys. Soc. Jpn}\ }\textbf {\bibinfo {volume} {67}},\ \bibinfo
  {pages} {1822} (\bibinfo {year} {1998})}\BibitemShut {NoStop}%
\bibitem [{\citenamefont {Chang}\ \emph {et~al.}(2004)\citenamefont {Chang},
  \citenamefont {Hamley}, \citenamefont {Barrett}, \citenamefont {Sauer},
  \citenamefont {Fortier}, \citenamefont {Zhang}, \citenamefont {You},\ and\
  \citenamefont {Chapman}}]{Chang2004a}%
  \BibitemOpen
  \bibfield  {author} {\bibinfo {author} {\bibfnamefont {M.-S.}\ \bibnamefont
  {Chang}}, \bibinfo {author} {\bibfnamefont {C.~D.}\ \bibnamefont {Hamley}},
  \bibinfo {author} {\bibfnamefont {M.~D.}\ \bibnamefont {Barrett}}, \bibinfo
  {author} {\bibfnamefont {J.~A.}\ \bibnamefont {Sauer}}, \bibinfo {author}
  {\bibfnamefont {K.~M.}\ \bibnamefont {Fortier}}, \bibinfo {author}
  {\bibfnamefont {W.}~\bibnamefont {Zhang}}, \bibinfo {author} {\bibfnamefont
  {L.}~\bibnamefont {You}}, \ and\ \bibinfo {author} {\bibfnamefont {M.~S.}\
  \bibnamefont {Chapman}},\ }\href {\doibase 10.1103/PhysRevLett.92.140403}
  {\bibfield  {journal} {\bibinfo  {journal} {Phys. Rev. Lett.}\ }\textbf
  {\bibinfo {volume} {92}},\ \bibinfo {pages} {140403} (\bibinfo {year}
  {2004})}\BibitemShut {NoStop}%
\bibitem [{\citenamefont {Gerbier}\ \emph {et~al.}(2006)\citenamefont
  {Gerbier}, \citenamefont {Widera}, \citenamefont {F\"olling}, \citenamefont
  {Mandel},\ and\ \citenamefont {Bloch}}]{Gerbier2006a}%
  \BibitemOpen
  \bibfield  {author} {\bibinfo {author} {\bibfnamefont {F.}~\bibnamefont
  {Gerbier}}, \bibinfo {author} {\bibfnamefont {A.}~\bibnamefont {Widera}},
  \bibinfo {author} {\bibfnamefont {S.}~\bibnamefont {F\"olling}}, \bibinfo
  {author} {\bibfnamefont {O.}~\bibnamefont {Mandel}}, \ and\ \bibinfo {author}
  {\bibfnamefont {I.}~\bibnamefont {Bloch}},\ }\href {\doibase
  10.1103/PhysRevA.73.041602} {\bibfield  {journal} {\bibinfo  {journal} {Phys.
  Rev. A}\ }\textbf {\bibinfo {volume} {73}},\ \bibinfo {pages} {041602}
  (\bibinfo {year} {2006})}\BibitemShut {NoStop}%
\bibitem [{\citenamefont {Kawaguchi}\ and\ \citenamefont
  {Ueda}(2012)}]{Kawaguchi2012R}%
  \BibitemOpen
  \bibfield  {author} {\bibinfo {author} {\bibfnamefont {Y.}~\bibnamefont
  {Kawaguchi}}\ and\ \bibinfo {author} {\bibfnamefont {M.}~\bibnamefont
  {Ueda}},\ }\href {\doibase http://dx.doi.org/10.1016/j.physrep.2012.07.005}
  {\bibfield  {journal} {\bibinfo  {journal} {Physics Reports}\ }\textbf
  {\bibinfo {volume} {520}},\ \bibinfo {pages} {253 } (\bibinfo {year}
  {2012})}\BibitemShut {NoStop}%
\bibitem [{\citenamefont {Saito}\ \emph {et~al.}(2007)\citenamefont {Saito},
  \citenamefont {Kawaguchi},\ and\ \citenamefont {Ueda}}]{Saito2007a}%
  \BibitemOpen
  \bibfield  {author} {\bibinfo {author} {\bibfnamefont {H.}~\bibnamefont
  {Saito}}, \bibinfo {author} {\bibfnamefont {Y.}~\bibnamefont {Kawaguchi}}, \
  and\ \bibinfo {author} {\bibfnamefont {M.}~\bibnamefont {Ueda}},\ }\href
  {\doibase 10.1103/PhysRevA.76.043613} {\bibfield  {journal} {\bibinfo
  {journal} {Phys. Rev. A}\ }\textbf {\bibinfo {volume} {76}},\ \bibinfo
  {pages} {043613} (\bibinfo {year} {2007})}\BibitemShut {NoStop}%
\bibitem [{\citenamefont {Barnett}\ \emph {et~al.}(2011)\citenamefont
  {Barnett}, \citenamefont {Polkovnikov},\ and\ \citenamefont
  {Vengalattore}}]{Barnett2011}%
  \BibitemOpen
  \bibfield  {author} {\bibinfo {author} {\bibfnamefont {R.}~\bibnamefont
  {Barnett}}, \bibinfo {author} {\bibfnamefont {A.}~\bibnamefont
  {Polkovnikov}}, \ and\ \bibinfo {author} {\bibfnamefont {M.}~\bibnamefont
  {Vengalattore}},\ }\href {\doibase 10.1103/PhysRevA.84.023606} {\bibfield
  {journal} {\bibinfo  {journal} {Phys. Rev. A}\ }\textbf {\bibinfo {volume}
  {84}},\ \bibinfo {pages} {023606} (\bibinfo {year} {2011})}\BibitemShut
  {NoStop}%
\bibitem [{\citenamefont {Zhang}\ \emph {et~al.}(2005)\citenamefont {Zhang},
  \citenamefont {Zhou}, \citenamefont {Chang}, \citenamefont {Chapman},\ and\
  \citenamefont {You}}]{zhang2005}%
  \BibitemOpen
  \bibfield  {author} {\bibinfo {author} {\bibfnamefont {W.}~\bibnamefont
  {Zhang}}, \bibinfo {author} {\bibfnamefont {D.~L.}\ \bibnamefont {Zhou}},
  \bibinfo {author} {\bibfnamefont {M.-S.}\ \bibnamefont {Chang}}, \bibinfo
  {author} {\bibfnamefont {M.~S.}\ \bibnamefont {Chapman}}, \ and\ \bibinfo
  {author} {\bibfnamefont {L.}~\bibnamefont {You}},\ }\href {\doibase
  10.1103/PhysRevLett.95.180403} {\bibfield  {journal} {\bibinfo  {journal}
  {Phys. Rev. Lett.}\ }\textbf {\bibinfo {volume} {95}},\ \bibinfo {pages}
  {180403} (\bibinfo {year} {2005})}\BibitemShut {NoStop}%
\bibitem [{\citenamefont {Lamacraft}(2008)}]{lamacraft2008}%
  \BibitemOpen
  \bibfield  {author} {\bibinfo {author} {\bibfnamefont {A.}~\bibnamefont
  {Lamacraft}},\ }\href {\doibase 10.1103/PhysRevA.77.063622} {\bibfield
  {journal} {\bibinfo  {journal} {Phys. Rev. A}\ }\textbf {\bibinfo {volume}
  {77}},\ \bibinfo {pages} {063622} (\bibinfo {year} {2008})}\BibitemShut
  {NoStop}%
\bibitem [{\citenamefont {T{\"a}uber}(2014)}]{Tauber2014}%
  \BibitemOpen
  \bibfield  {author} {\bibinfo {author} {\bibfnamefont {U.}~\bibnamefont
  {T{\"a}uber}},\ }\href@noop {} {\emph {\bibinfo {title} {Critical
  dynamics}}}\ (\bibinfo  {publisher} {Cambridge University Press},\ \bibinfo
  {year} {2014})\BibitemShut {NoStop}%
\bibitem [{\citenamefont {Kudo}(2016)}]{kudo2016}%
  \BibitemOpen
  \bibfield  {author} {\bibinfo {author} {\bibfnamefont {K.}~\bibnamefont
  {Kudo}},\ }\href {\doibase 10.1103/PhysRevE.94.062215} {\bibfield  {journal}
  {\bibinfo  {journal} {Phys. Rev. E}\ }\textbf {\bibinfo {volume} {94}},\
  \bibinfo {pages} {062215} (\bibinfo {year} {2016})}\BibitemShut {NoStop}%
\bibitem [{\citenamefont {Su}\ \emph {et~al.}(2011)\citenamefont {Su},
  \citenamefont {Hsueh}, \citenamefont {Liu}, \citenamefont {Horng},
  \citenamefont {Tsai}, \citenamefont {Gou},\ and\ \citenamefont
  {Liu}}]{Su2011a}%
  \BibitemOpen
  \bibfield  {author} {\bibinfo {author} {\bibfnamefont {S.-W.}\ \bibnamefont
  {Su}}, \bibinfo {author} {\bibfnamefont {C.-H.}\ \bibnamefont {Hsueh}},
  \bibinfo {author} {\bibfnamefont {I.-K.}\ \bibnamefont {Liu}}, \bibinfo
  {author} {\bibfnamefont {T.-L.}\ \bibnamefont {Horng}}, \bibinfo {author}
  {\bibfnamefont {Y.-C.}\ \bibnamefont {Tsai}}, \bibinfo {author}
  {\bibfnamefont {S.-C.}\ \bibnamefont {Gou}}, \ and\ \bibinfo {author}
  {\bibfnamefont {W.~M.}\ \bibnamefont {Liu}},\ }\href {\doibase
  10.1103/PhysRevA.84.023601} {\bibfield  {journal} {\bibinfo  {journal} {Phys.
  Rev. A}\ }\textbf {\bibinfo {volume} {84}},\ \bibinfo {pages} {023601}
  (\bibinfo {year} {2011})}\BibitemShut {NoStop}%
\bibitem [{\citenamefont {Rooney}\ \emph {et~al.}(2012)\citenamefont {Rooney},
  \citenamefont {Blakie},\ and\ \citenamefont {Bradley}}]{Rooney2012a}%
  \BibitemOpen
  \bibfield  {author} {\bibinfo {author} {\bibfnamefont {S.~J.}\ \bibnamefont
  {Rooney}}, \bibinfo {author} {\bibfnamefont {P.~B.}\ \bibnamefont {Blakie}},
  \ and\ \bibinfo {author} {\bibfnamefont {A.~S.}\ \bibnamefont {Bradley}},\
  }\href {\doibase 10.1103/PhysRevA.86.053634} {\bibfield  {journal} {\bibinfo
  {journal} {Phys. Rev. A}\ }\textbf {\bibinfo {volume} {86}},\ \bibinfo
  {pages} {053634} (\bibinfo {year} {2012})}\BibitemShut {NoStop}%
\bibitem [{\citenamefont {Bradley}\ and\ \citenamefont
  {Blakie}(2014)}]{Bradley2014a}%
  \BibitemOpen
  \bibfield  {author} {\bibinfo {author} {\bibfnamefont {A.~S.}\ \bibnamefont
  {Bradley}}\ and\ \bibinfo {author} {\bibfnamefont {P.~B.}\ \bibnamefont
  {Blakie}},\ }\href {\doibase 10.1103/PhysRevA.90.023631} {\bibfield
  {journal} {\bibinfo  {journal} {Phys. Rev. A}\ }\textbf {\bibinfo {volume}
  {90}},\ \bibinfo {pages} {023631} (\bibinfo {year} {2014})}\BibitemShut
  {NoStop}%
\bibitem [{\citenamefont {Vinit}\ and\ \citenamefont
  {Raman}(2017)}]{Vinit2017a}%
  \BibitemOpen
  \bibfield  {author} {\bibinfo {author} {\bibfnamefont {A.}~\bibnamefont
  {Vinit}}\ and\ \bibinfo {author} {\bibfnamefont {C.}~\bibnamefont {Raman}},\
  }\href {\doibase 10.1103/PhysRevA.95.011603} {\bibfield  {journal} {\bibinfo
  {journal} {Phys. Rev. A}\ }\textbf {\bibinfo {volume} {95}},\ \bibinfo
  {pages} {011603} (\bibinfo {year} {2017})}\BibitemShut {NoStop}%
\bibitem [{\citenamefont {Furukawa}(1985)}]{furukawa1985}%
  \BibitemOpen
  \bibfield  {author} {\bibinfo {author} {\bibfnamefont {H.}~\bibnamefont
  {Furukawa}},\ }\href {\doibase 10.1103/PhysRevA.31.1103} {\bibfield
  {journal} {\bibinfo  {journal} {Phys. Rev. A}\ }\textbf {\bibinfo {volume}
  {31}},\ \bibinfo {pages} {1103} (\bibinfo {year} {1985})}\BibitemShut
  {NoStop}%
\bibitem [{\citenamefont {Nowak}\ \emph {et~al.}(2012)\citenamefont {Nowak},
  \citenamefont {Erne}, \citenamefont {Karl}, \citenamefont {Schole},
  \citenamefont {Sexty},\ and\ \citenamefont {Gasenzer}}]{nowak2012}%
  \BibitemOpen
  \bibfield  {author} {\bibinfo {author} {\bibfnamefont {B.}~\bibnamefont
  {Nowak}}, \bibinfo {author} {\bibfnamefont {S.}~\bibnamefont {Erne}},
  \bibinfo {author} {\bibfnamefont {M.}~\bibnamefont {Karl}}, \bibinfo {author}
  {\bibfnamefont {J.}~\bibnamefont {Schole}}, \bibinfo {author} {\bibfnamefont
  {D.}~\bibnamefont {Sexty}}, \ and\ \bibinfo {author} {\bibfnamefont
  {T.}~\bibnamefont {Gasenzer}},\ }in\ \href {\doibase
  10.1093/acprof:oso/9780198768166.003.0007} {\emph {\bibinfo {booktitle}
  {Strongly Interacting Quantum Systems out of Equilibrium: Lecture Notes of
  the Les Houches Summer School}}},\ Vol.~\bibinfo {volume} {99}\ (\bibinfo
  {publisher} {Oxford University Press},\ \bibinfo {year} {2012})\
  Chap.~\bibinfo {chapter} {7}\BibitemShut {NoStop}%
\bibitem [{\citenamefont {Langen}\ \emph {et~al.}(2016)\citenamefont {Langen},
  \citenamefont {Gasenzer},\ and\ \citenamefont {Schmiedmayer}}]{langen2016}%
  \BibitemOpen
  \bibfield  {author} {\bibinfo {author} {\bibfnamefont {T.}~\bibnamefont
  {Langen}}, \bibinfo {author} {\bibfnamefont {T.}~\bibnamefont {Gasenzer}}, \
  and\ \bibinfo {author} {\bibfnamefont {J.}~\bibnamefont {Schmiedmayer}},\
  }\href {https://doi.org/10.1088/1742-5468/2016/06/064009} {\bibfield
  {journal} {\bibinfo  {journal} {J. Stat. Mech.}\ }\textbf {\bibinfo {volume}
  {2016}},\ \bibinfo {pages} {064009} (\bibinfo {year} {2016})}\BibitemShut
  {NoStop}%
\bibitem [{\citenamefont {Karl}\ \emph {et~al.}(2013)\citenamefont {Karl},
  \citenamefont {Nowak},\ and\ \citenamefont {Gasenzer}}]{karl2013}%
  \BibitemOpen
  \bibfield  {author} {\bibinfo {author} {\bibfnamefont {M.}~\bibnamefont
  {Karl}}, \bibinfo {author} {\bibfnamefont {B.}~\bibnamefont {Nowak}}, \ and\
  \bibinfo {author} {\bibfnamefont {T.}~\bibnamefont {Gasenzer}},\ }\href
  {\doibase 10.1103/PhysRevA.88.063615} {\bibfield  {journal} {\bibinfo
  {journal} {Phys. Rev. A}\ }\textbf {\bibinfo {volume} {88}},\ \bibinfo
  {pages} {063615} (\bibinfo {year} {2013})}\BibitemShut {NoStop}%
\bibitem [{\citenamefont {Mukerjee}\ \emph {et~al.}(2006)\citenamefont
  {Mukerjee}, \citenamefont {Xu},\ and\ \citenamefont {Moore}}]{mukerjee2006}%
  \BibitemOpen
  \bibfield  {author} {\bibinfo {author} {\bibfnamefont {S.}~\bibnamefont
  {Mukerjee}}, \bibinfo {author} {\bibfnamefont {C.}~\bibnamefont {Xu}}, \ and\
  \bibinfo {author} {\bibfnamefont {J.~E.}\ \bibnamefont {Moore}},\ }\href
  {\doibase 10.1103/PhysRevLett.97.120406} {\bibfield  {journal} {\bibinfo
  {journal} {Phys. Rev. Lett.}\ }\textbf {\bibinfo {volume} {97}},\ \bibinfo
  {pages} {120406} (\bibinfo {year} {2006})}\BibitemShut {NoStop}%
\bibitem [{\citenamefont {Williamson}\ and\ \citenamefont
  {Blakie}(2016{\natexlab{c}})}]{williamson2016c}%
  \BibitemOpen
  \bibfield  {author} {\bibinfo {author} {\bibfnamefont {L.~A.}\ \bibnamefont
  {Williamson}}\ and\ \bibinfo {author} {\bibfnamefont {P.~B.}\ \bibnamefont
  {Blakie}},\ }\href {\doibase 10.1103/PhysRevA.94.063615} {\bibfield
  {journal} {\bibinfo  {journal} {Phys. Rev. A}\ }\textbf {\bibinfo {volume}
  {94}},\ \bibinfo {pages} {063615} (\bibinfo {year}
  {2016}{\natexlab{c}})}\BibitemShut {NoStop}%
\bibitem [{\citenamefont {Kasamatsu}\ \emph {et~al.}(2016)\citenamefont
  {Kasamatsu}, \citenamefont {Eto},\ and\ \citenamefont
  {Nitta}}]{kasamatsu2016}%
  \BibitemOpen
  \bibfield  {author} {\bibinfo {author} {\bibfnamefont {K.}~\bibnamefont
  {Kasamatsu}}, \bibinfo {author} {\bibfnamefont {M.}~\bibnamefont {Eto}}, \
  and\ \bibinfo {author} {\bibfnamefont {M.}~\bibnamefont {Nitta}},\ }\href
  {\doibase 10.1103/PhysRevA.93.013615} {\bibfield  {journal} {\bibinfo
  {journal} {Phys. Rev. A}\ }\textbf {\bibinfo {volume} {93}},\ \bibinfo
  {pages} {013615} (\bibinfo {year} {2016})}\BibitemShut {NoStop}%
\end{thebibliography}

%

\end{document}